\documentclass[prd,aps]{revtex4}

\usepackage{epsfig}
\newcommand{\beq}{\begin{equation}}
\newcommand{\eeq}{\end{equation}}
\newcommand{\be}{\begin{eqnarray}}
\newcommand{\ee}{\end{eqnarray}}

\newcommand{\bs}{B_s-\bar{B}_s}
\newcommand{\bd}{B_d-\bar{B}_d}
\newcommand{\bq}{B_q-\bar{B}_q}

\begin{document}

%\draft
%\preprint{
%\begin{tabular}{l}
%\hbox to\hsize{June, 2008\hfill KIAS-P08040}\\[-2mm]
%\hbox to\hsize{              \hfill KAIST-TH 02/04}\\[-3mm]
%\hbox to\hsize{           \hfill hep-ph/yymmdd}\\[5mm] 
%\end{tabular}
%}

\title{
A like-sign dimuon charge asymmetry at Tevatron 
\\
induced by the anomalous top quark couplings
}

\author{ Jong Phil Lee }
%\email{jplee@kias.re.kr}

\affiliation{
Department of Physics and IPAP, Yonsei University, Seoul 120-749, Korea\\
Division of Quantum Phases \& Devices,
School of Physics, Konkuk University, Seoul 143-701, Korea
}

\author{ Kang Young Lee }
\email{kylee14214@gmail.com}

\affiliation{
Division of Quantum Phases \& Devices,
School of Physics, Konkuk University, Seoul 143-701, Korea
}

\date{\today}
%\maketitle

\begin{abstract}

We show that the recently measured 
3.9 $\sigma$ deviations of the charge asymmetry of like-sign dimuon events 
from the standard model prediction
by the D0 collaboration at Tevatron
can be explained by introducing the anomalous 
right-handed top quark couplings.
Combined analysis with the $\bs$, $\bd$ mixings, 
$B \to X_s \gamma$ decays and the time-dependent CP asymmetry
in $B \to \phi K$ decays has been performed.
The anomalous $tsW$ couplings are preferred 
to explain the dimuon charge asymmetry
by other CP violating observables.

\end{abstract}

\maketitle

%\tightenlines

               %%%%%%%%%%%%%%%%%%%%%%%%%%%%%%%
               %%    BEGINNING OF TEXT      %%
               %%%%%%%%%%%%%%%%%%%%%%%%%%%%%%%

\section{Introduction}

Recently the D0 collaboration has measured the CP violating
like-sign dimuon charge asymmetry for $b$ hadrons, defined as
\be
A_{sl}^b \equiv \frac{N_b^{++} - N_b^{--}}{N_b^{++} + N_b^{--}},
\ee
of which value is reported to be
\cite{d0new}  
\be
A_{sl}^b = (-0.957 \pm 0.172~({\rm stat.}) \pm 0.093~({\rm syst.})) \%,
\ee
in an integrated luminosity of 9.0 fb$^{-1}$ of $p \bar{p}$ data 
at $\sqrt{s}=1.96$ TeV at Tevatron.
In the definition of Eq. (1), 
$N_b^{++}$ and $N_b^{--}$ are the number of events
where two $b$ hadrons semileptonically decay into muons
with charges of the same sign.
Since the $b$ quarks are produced as $b \bar{b}$ pairs 
from $p \bar{p}$ collisions at Tevatron,
the like-sign dimuon events arises from a direct semileptonic decay
of one of $b$ hadrons and a semileptonic decay of the other $b$ hadron
following the $B^0-\bar{B}^0$ oscillation.
In the standard model (SM), the source of the CP violation
in the neutral $B^0_q$ system is the phase of 
the Cabibbo-Kobayashi-Maskawa (CKM) matrix elements involved 
in the box diagram.
The D0 measurement of Eq. (2) shows a deviation of
3.9 $\sigma$ from the SM prediction
$A_{sl}^b = (-2.3^{+0.5}_{-0.6}) \times 10^{-4}$.
The measured value of Eq. (2) is improved again by more data 
as more data is analyzed.
If the deviation is confirmed with other experiments,
it indicates the existence of the new physics beyond the SM.
Many works are devoted to explanation
of the D0 dimuon asymmetry in and beyond the SM
\cite{dimuon}.

Although the charged currents are purely left-handed in the SM,
the existence of right-handed charged currents 
is predicted in many new physics models beyond the SM.
For instance, 
the variant SU(2)$_L \times$SU(2)$_R \times$U(1) model \cite{lr} 
and a dynamical electroweak symmetry breaking model \cite{dsb}
predicts additional right-handed currents and 
some modification of the left-handed currents.
In this work, 
we study the effects of the anomalous right-handed top quark couplings
on the D0 like-sign dimuon charge asymmetry.
We introduce additional right-handed top quark couplings
without specifying the underlying model
and assume no effects of new particles and 
additional neutral currents interactions.
Impacts of the anomalous top quark couplings 
have been studied in flavour physics and at colliders 
\cite{lee1,lee2,anomalous}.
Here, we show that the measurement of the $A_{sl}^b$ 
can be explained by 
both of the anomalous $tsW$ and $tbW$ couplings,
with accommodation of present data of Br$(B \to X_s \gamma)$,
$\Delta M_s$, $\Delta M_d$ 
and CP asymmetry in $B \to \phi K$ decays
at 2-$\sigma$ level.

This paper is organized as follows.
In section II, we present the formalism for the dimuon charge asymmetry
and neutral $B$ meson system.
In section III, we present the contribution of 
the anomalous top quark couplings to $B \to X_s \gamma$,
$B - \bar{B}$ mixings, and $B \to \phi K$ decays 
to obtain the possible parameter sets.
In section IV, we discuss the dimuon charge symmetry
with the anomalous top quark couplings
and future experiments.
Finally we conclude in section V.

\section{Dimuon charge asymmetry in the neutral $B$ meson system}

Since the like-sign dimuon events following $b \bar{b}$ production 
arise through the $B-\bar{B}$ oscillation,
the dimuon charge asymmetry can be described in terms of
the parameters of the $B-\bar{B}$ mixings.
The neutral $B$ meson system is described by the Schr\"odinger equation
\be
i \frac{d}{dt} \left(
\begin{array}{c}
B_q(t) \\
\bar{B}_q(t) \\
\end{array}
\right)
= \left( M - \frac{i}{2} \Gamma \right)
\left(
\begin{array}{c}
B_q(t) \\
\bar{B}_q(t) \\
\end{array}
\right),
\ee
where $M$ is the mass matrix and $\Gamma$ the decay matrix with $q=d,s$.
The $\Delta B = 2$ transition amplitudes 
\be
\langle B_q^0 | {\cal H}_{\rm eff}^{\Delta B = 2} | \bar{B}_q^0 \rangle
         = M_{12}^q,
\ee
leads to the mass difference between the heavy and the light states
of $B$ meson,
\be
\Delta M_q \equiv M_H^{q} - M_L^{q} = 2 | M_{12}^q |,
\ee
where $M_H^{B_q}$ and $M_L^{B_q}$
are the mass eigenvalues for the heavy and the light eigenstates
respectively.
The total decay width difference of the mass eigenstates is defined by
\be
\Delta \Gamma_q \equiv \Gamma_L^q - \Gamma_H^q
                = 2 |\Gamma_{12}^q| \cos \phi_q,
\ee
where the decay widths $\Gamma_L$ and $\Gamma_H$ are corresponding to
the physical eigenstates $B_L$ and $B_H$ respectively
and the CP phase is 
$\phi_q \equiv {\rm arg}\left( - M_{12}^q/\Gamma_{12}^q \right)$.

The like-sign dimuon events consist of a right-sign (RS) process 
and a wrong-sign (WS) process,
\be
A_{sl}^b \equiv \frac{ \Gamma(b \bar{b} \to \mu^+ \mu^+ X)
                       - \Gamma(b \bar{b} \to \mu^- \mu^- X)}
                       { \Gamma(b \bar{b} \to \mu^+ \mu^+ X)
                       + \Gamma(b \bar{b} \to \mu^- \mu^- X)}
= \frac{ \Gamma_{RS}^+ \Gamma_{WS}^+ - \Gamma_{RS}^- \Gamma_{WS}^-}
         { \Gamma_{RS}^+ \Gamma_{WS}^+ + \Gamma_{RS}^- \Gamma_{WS}^-},
\ee
in which $\Gamma_{RS}$ denotes the direct semileptonic decay rate 
in the right-sign process 
and $\Gamma_{WS}$ the decay in the wrong-sign process 
implying the semileptonic decay rate of 
the $B_q^0 (\bar{B}_q^0)$ meson following 
$B_q^0-\bar{B}_q^0$ oscillation.
The dimuon asymmetry implies the CP violation in the $B$ system.
%we have
%\be
%A_{sl}^b = \frac{f_d T_d^- + f_s T_s^-}{f_d T_d^+ + f_s T_s^+},
%\ee
%where $f_q$ is the production fraction of $B^0_q$ and
%$T_q^\pm$ are the time integrated probability defined by
%$T_q^\pm = T(\bar{B}_q \to B_q) \pm T(B_q \to \bar{B}_q)$.

The asymmetry of dimuon events is derived from the charge asymmetry
of semileptonic decays of neutral $B^0_q$ mesons, $a_{sl}^q$
defined as
\be
a_{sl}^q \equiv \frac{\Gamma(\bar{B}^0_q(t) \to \mu^+ X)
                     - \Gamma(B^0_q(t) \to \mu^- X)}
                     {\Gamma(\bar{B}^0_q(t) \to \mu^+ X)
                     + \Gamma(B^0_q(t) \to \mu^- X)}.
\ee
At Tevatron experiment, 
both decays of $B_d$ and $B_s$ mesons contribute to the asymmetry.
Assuming that $\Gamma(B_d^0 \to \mu^+ X)=\Gamma(B_s^0 \to \mu^+ X)$
to a very good approximation, 
the like-sign dimuon charge asymmetry can be expressed
in terms of $a_{sl}^q$ as \cite{grossman} 
\be
A_{sl}^b = \frac{1}{f_d Z_d + f_s Z_s}
           \left( f_d Z_d a_{sl}^d + f_s Z_s a_{sl}^s \right)
\ee
where $f_q$ are the production fractions of $B_q$ mesons, and 
$Z_q = 1/(1-y_q^2)-1/(1+x_q^2)$
with $y_q = \Delta \Gamma_q/(2 \Gamma_q)$, $x_q=\Delta M_q/\Gamma_q$.
These parameters are measured to be 
$f_d = 0.323 \pm 0.037$,
$f_s = 0.118 \pm 0.015$,
$x_d = 0.774 \pm 0.008$,
$x_s = 26.2 \pm 0.5$,
and
$y_d = 0$,
$y_s = 0.046 \pm 0.027$
\cite{pdg}.
With these values, Eq. (10) is rewritten by
\be
A_{sl}^b = (0.506 \pm 0.043) a_{sl}^d + (0.494 \pm 0.043) a_{sl}^s.
\ee

The charge asymmetry for wrong charge semileptonic decay 
in Eq. (9) is expressed as
\be
a_{sl}^q = \frac{|\Gamma_{12}^q|}{|M_{12}^q|} \sin \phi_q
         = \frac{\Delta \Gamma_q}{\Delta M_q} \tan \phi_q,
\ee
of which the SM predictions are given by
\cite{nierste}
\be
a_{sl}^{d} = (-4.8^{+1.0}_{-1.2}) \times 10^{-4},
\nonumber \\
a_{sl}^{s} = (2.1 \pm 0.6) \times 10^{-5}.
\ee

In the SM, $\Delta \Gamma_d/\Gamma_d$ is less than $1 \%$,
while $\Delta \Gamma_s/\Gamma_s \sim 10 \%$ is rather large. 
The decay matrix elements $\Gamma_{12}^q$ is obtained from
the tree level decays $b \to c \bar{c} q$.
Since the anomalous top couplings affects $\Gamma_{12}^q$
through loops only,
we ignore the new physics effects on $\Gamma_{12}^q$ in this work.

\section{Anomalous top quark couplings and $B$ physics}

In this paper, we work with an effective Lagrangian 
in a model independent way
to parameterize the new physics effects.
After fixing the phases of quarks 
so that $V_{tq}^{\rm SM}$ are the CKM matrix elements of the SM,
we introduce the new $Wtq$ couplings $g_L^q$ and $g_R^q$
to redefine the effective CKM matrix elements and
right-handed couplings:
\be
{\cal L} &=& -\frac{g}{\sqrt{2}} \sum_{q=s,b}
          V_{tq}^{\rm SM}~ 
          \left( \bar{t} \gamma^\mu P_L q W^+_\mu
             + \bar{t} \gamma^\mu (g_L^q P_L + g_R^q P_R) q W^+_\mu \right)
  + H.c.,
\nonumber \\
         &=& -\frac{g}{\sqrt{2}} \sum_{q=s,b}
          V_{tq}^{\rm eff}~ \bar{t} \gamma^\mu
                                 (P_L + \xi_q P_R) q W^+_\mu
  + H.c.,
\ee
where 
$V_{tq}^{\rm eff} = V_{tq}^{\rm SM} (1+g_L^q)$, 
and $V_{tq}^{\rm eff} \xi_q = V_{tq}^{\rm SM} g_R^q$. 
Since we set $g_L^q$ and $g_R^q$ to be complex,
$V_{tq}^{\rm eff}$ and $\xi_q$ involve new phases
and will predict new CP violating processes in $B$ physics.
For simplicity, 
we assume that either one of anomalous $tsW$ or $tbW$ couplings 
is nonzero in this analysis.
Then other CKM matrix elements are same as those in the SM
and the phase of quarks are fixed with them.

The matrix elements of the third row of the CKM matrix
are not directly measured yet,
but just indirectly constrained by loop-induced processes
and the unitarity of the CKM matrix.
In our framework, the constraints should be applied to 
effective CKM matrix elements $V_{tq}^{\rm eff}$
instead of $V_{tq}^{\rm SM}$.
The additional $V_{tq}^{\rm eff} \xi_q$ terms measure 
the anomalous right-handed top couplings. 
Effects on $W \bar{t} d$ coupling are ignored here
due to the smallness of $V_{td}$.

\subsection{$B \to X_s \gamma$ decays}

Contributions of the right-handed top quark couplings to
the penguin diagram for $b \to s$ transition are enhanced
by the factor of $m_t/m_b$.
Thus the radiative $B \to X_s \gamma$ decays are sensitive
to the anomalous right-handed $W\bar{t}b$ and $W\bar{t}s$ couplings
and provides strong constraints on them.

The $\Delta B =1$ effective Hamiltonian for $B \to X_s \gamma$ process 
with the right-handed couplings is given by 
\be
{\cal H}_{eff}^{\Delta B=1} = -\frac{4 G_F}{\sqrt{2}} V_{ts}^* V_{tb} 
           \sum_{i=1}^{8}
             \left( C_i(\mu) O_i(\mu) + C'_i(\mu) O'_i(\mu) \right),
\ee
where the dimension 6 operators $O_i$ are given in the Ref. \cite{buras},
and $O'_i$ are their chiral conjugate operators.
The SM Wilson coefficients are shifted by
$C_7(m_W) = F(x_t) + \xi_b (m_t/m_b) F_R(x_t)$ and
$C_8(m_W) = G(x_t) + \xi_b (m_t/m_b) G_R(x_t)$
while the new Wilson coefficients are formed as 
$C'_7(m_W) = \xi_s (m_t/m_b) F_R(x_t)$ and
$C'_8(m_W) = \xi_s (m_t/m_b) G_R(x_t)$
%\be
%C_2(m_W) &=& -1,
%\nonumber \\
%C_7(m_W) &=& F(x_t) + \xi_b \frac{m_t}{m_b} F_R(x_t),
%\nonumber \\
%C_8(m_W) &=& G(x_t) + \xi_b \frac{m_t}{m_b} G_R(x_t),
%\nonumber \\
%C'_7(m_W) &=& \xi_s \frac{m_t}{m_b} F_R(x_t),
%\nonumber \\
%C'_8(m_W) &=& \xi_s \frac{m_t}{m_b} G_R(x_t),
%\nonumber \\
%C_i(m_W) &=& C'_i(m_W) = 0, ~~~~~~{\rm otherwise,}
%\ee
in the leading order of $\xi_q$.
The Inami-Lim loop functions 
$F(x)$ and $G(x)$ are given by in Ref. \cite{buras,inami}
and the new loop functions 
$F_R(x)$ and $G_R(x)$ can be found in Ref. \cite{cho,lee1,lee2}.

The branching ratio of the $B \to X_s \gamma$ decays 
including $\xi_s$ and $\xi_b$ effects is given by
\be
{\rm Br}(B \to X_s \gamma) &=& {\rm Br}^{\rm SM}(B \to X_s \gamma)
     \left( \frac{|{V_{ts}^{\rm eff}}^* V_{tb}^{\rm eff}|}
                 {0.0404} \right)^2
     \left[ 1 + Re (\xi_b) \frac{m_t}{m_b} \left(
       0.68 \frac{F_R(x_t)}{F(x_t)} + 0.07 \frac{G_R(x_t)}{G(x_t)} \right)
     \right.
\nonumber \\
     &&~~~~~~~~~~~~   
     \left.
         + ( |\xi_b|^2 + |\xi_s|^2 ) \frac{m_t^2}{m_b^2} 
         \left( 0.112 \frac{F^2_R(x_t)}{F^2(x_t)} 
              + 0.002 \frac{G^2_R(x_t)}{G^2(x_t)} 
              + 0.025 \frac{F_R(x_t) G_R(x_t)}{F(x_t) G(x_t)} \right)
     \right],
\ee
where the numerical values are obtained by the RG evolution 
in Ref. \cite{kagan}.
The SM prediction of the branching ratio is given by
\cite{bsgammaSM}
$ {\rm Br}(B \to X_s \gamma) = (3.15 \pm 0.23) \times 10^{-4} $
and the current world average value
of the measured branching ratio given by
\cite{hfag}
$ {\rm Br}(B \to X_s \gamma) = (3.55 \pm 0.24 ^{+0.09}_{-0.10} \pm 0.03) 
                             \times 10^{-4} $
with the photon energy cut $E_\gamma > 1.6$ GeV. 

\begin{figure}[t!]
\centering
%\hbox to\textwidth{\hss\epsfig{file=fig1.eps,width=10cm,height=10cm}\hss}
% \vskip -0.5cm
\includegraphics[width=10cm]{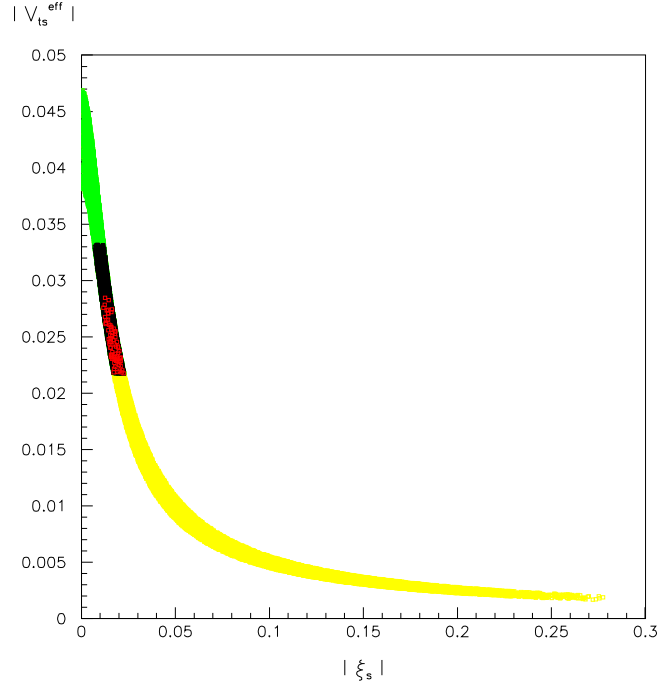}
\caption{ \small
Allowed parameters $(|\xi_s|,|V_{ts}^{\rm eff}|)$ under the $B$ physics 
constraints and D0 dimuon asymmetry.
The whole band of the green (grey) + black + yellow (light grey) 
regions is allowed by Br$(B \to X_s \gamma)$ only.
The green (grey) + black regions are allowed by
Br$(B \to X_s \gamma)$ and $\Delta M_s$.
The black region is allowed by both constraints of
Br$(B \to X_s \gamma)$ and $\Delta M_s$, 
and satisfies $A_{sl}^b$ measured by D0.
The red (dark grey) dots denote points additionally allowed by
CP asymmetries in $B \to \phi K$ decays.
The confidence level is at 95 \% C.L..
}
\end{figure}
\begin{figure}[t!]
\centering
%\hbox to\textwidth{\hss\epsfig{file=fig2.eps,width=10cm,height=10cm}\hss}
% \vskip -0.5cm
\includegraphics[width=10cm]{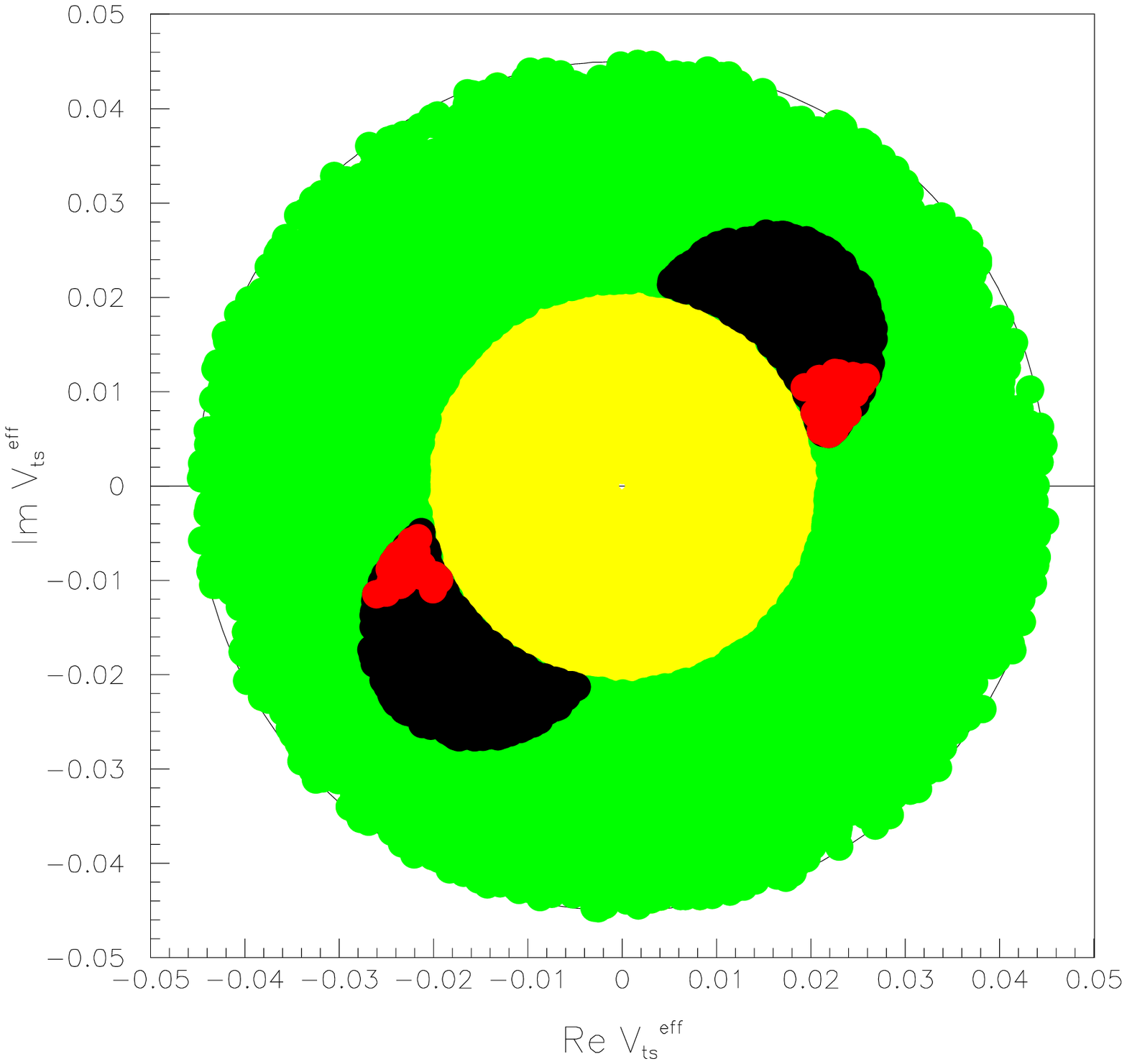}
\caption{\small
Allowed parameters $({\rm Re} V_{ts}^{\rm eff}, {\rm Im} V_{ts}^{\rm eff})$ 
under the $B$ physics constraints and D0 dimuon asymmetry.
The whole circle of the yellow (light grey) + green (grey) 
+ black regions is allowed by Br$(B \to X_s \gamma)$ only,
the ring shape of the green (grey) + black regions allowed 
by Br$(B \to X_s \gamma)$ and $\Delta M_s$.
The black regions allowed by both constraints of
Br$(B \to X_s \gamma)$ and $\Delta M_s$, 
and satisfies $A_{sl}^b$ measured by D0.
The red (dark grey) dots denote points additionally allowed by
CP asymmetries in $B \to \phi K$ decays.
The confidence level is at 95 \% C.L..
}
\end{figure}

\subsection{$B - \bar{B}$ mixings}

The transition amplitude $M_{12}^q$ for $\bq$ mixing
is obtained from the box diagrams in the SM. 
In our model, the top quark couplings in the box diagram 
is modified to include the right-handed couplings.
Since the loop integral including
the odd number of right-handed couplings vanishes,
the leading contribution of $\xi_q$ to $M_{12}$ is of quadratic order.
We write $M_{12}^{s,d}$ as
\be
M_{12}^s &=& M_{12}^{s,SM} 
%\frac{G_F^2 m_W^2}{12 \pi} m_{B_s} \eta_{B} \hat{B}_{B_s} f_{B_s}^2 S_0(x_t)
     \left( \frac{{V_{ts}^{\rm eff}}^* V_{tb}^{\rm eff}}{0.0404} \right)^2
%                       (\bar{b} \gamma^\mu P_L s) (\bar{b} \gamma_\mu P_L s)
\left( 1 + \frac{S_3(x_t)}{S_0(x_t)}
 \left(
 \frac{\xi_s^2}{4} 
  \frac{\langle \bar{B}_s^0 |(\bar{b} P_R s) (\bar{b} P_R s)|B_s^0 \rangle}
         {\langle \bar{B}_s^0 |(\bar{b} \gamma^\mu P_L s) 
                         (\bar{b} \gamma_\mu P_L s)|B_s^0 \rangle}
     \right.
     \right.
\nonumber \\
&&~~~~~~~~~~
\left.
\left.
+\frac{\xi_b^\ast \xi_s}{2} 
 \frac{\langle \bar{B}_s^0 |(\bar{b} P_L s) (\bar{b} P_R s)|B_s^0 \rangle}
         {\langle \bar{B}_s^0 |(\bar{b} \gamma^\mu P_L s)
                         (\bar{b} \gamma_\mu P_L s)|B_s^0 \rangle}
+\frac{{\xi_b^\ast}^2}{4} 
 \frac{\langle \bar{B}_s^0 |(\bar{b} P_L s) (\bar{b} P_L s)|B_s^0 \rangle}
         {\langle \bar{B}_s^0 |(\bar{b} \gamma^\mu P_L s) 
                         (\bar{b} \gamma_\mu P_L s)|B_s^0 \rangle}
                             \right)
\right),
%\nonumber \\
%&\equiv& M_{12}^{\rm SM,s} \cdot \Delta_s,
\ee
and
\be
M_{12}^d &=& M_{12}^{d,SM} 
%\frac{G_F^2 m_W^2}{12 \pi} m_{B_q}
%                 \eta_{q} \hat{B}_{B_q} f_{B_q}^2 S_0(x_t)
     \left(V_{tb}^{\rm eff} \right)^2
      \left( 1 + \frac{S_3(x_t)}{S_0(x_t)} \frac{{\xi_b^\ast}^2}{4}
 \frac{\langle B_d^0 |(\bar{b} P_L d) (\bar{b} P_L d)|\bar{B}_d^0 \rangle}
         {\langle B_d^0 |(\bar{b} \gamma^\mu P_L d)
                         (\bar{b} \gamma_\mu P_L d)|\bar{B}_d^0 \rangle}
\right),
\ee
where the Inami-Lim loop functions for new box diagrams are given by
\be
%S_0(x) &=& \frac{4x-11 x^2 + x^3}{4 (1-x)^2} - \frac{3 x^3}{2 (1-x)^3} \log x,
%\nonumber \\
S_3(x) &=& 4 x^2 \left( \frac{2}{(1-x)^2} + \frac{1+x}{(1-x)^3} \log x \right),
\ee
and the SM loop function $S_0(x)$ can be found elsewhere \cite{buras,inami}.
The hadronic matrix elements for the four quark operators 
are parameterized by
\cite{beneke}
\be
\langle \bar{B}_q^0 |(\bar{b} \gamma^\mu P_L q) 
                         (\bar{b} \gamma_\mu P_L q)|B_q^0 \rangle
&=& \frac{8}{3} f_{B_q}^2 \hat{B}_{B_q} m_{B_q}^2,
\nonumber \\
\langle \bar{B}_q^0 |(\bar{b} P_L q) (\bar{b} P_L q)|B_q^0 \rangle
&=& \langle \bar{B}_q^0 |(\bar{b} P_R q) (\bar{b} P_R q)|B_q^0 \rangle
= -\frac{5}{3} f_{B_q}^2 \hat{B}_{B_q} m_{B_q}^2 
                 \left( \frac{m_{B_q}}{m_b+m_q} \right)^2,
\nonumber \\
\langle \bar{B}_q^0 |(\bar{b} P_L q) (\bar{b} P_R q)|B_q^0 \rangle
&=& \frac{7}{3} f_{B_q}^2 m_{B_q}^2 \frac{m_q}{m_b},
\ee
where $\hat{B}_{B_q}$ is the Bag parameter and $f_{B_q}^2$ the decay constant.

The SM predictions of the mass differences are 
$\Delta M_d = 0.53 \pm 0.02 $ ps$^{-1}$ and
$\Delta M_s = 19.30 \pm 6.74 \pm 0.07$ ps$^{-1}$ \cite{nierste}.
and the measurements are
$\Delta M_d = 0.509 \pm 0.006 $ ps$^{-1}$ \cite{hfag} and
$\Delta M_s = 17.77 \pm 0.10 \pm 0.07$ ps$^{-1}$ \cite{nierste}.

\subsection{CP asymmetries in $B \to \phi K$ decays}

The $b \to s \bar{s} s$ transition responsible for
the $B \to \phi K$ decays arises at one-loop level in the SM,
where the gluon penguin contribution dominates.
Since $V_{ts}^{\rm SM}$ involves no complex phase 
in the leading order in the SM,
the weak phase $\sin 2 \beta$ measured in $B \to \phi K$ decays
should agree with that of $B \to J/\psi K$ decays
and the direct CP asymmetry of $B \to \phi K$ decays should vanish 
up to small pollution.

The decay amplitude of $B \to \phi K$ decays 
with anomalous top couplings
are given in Ref. \cite{lee2}.
We define the parameter $\lambda$ as
\be
\lambda = \sqrt{ \frac{M_{12}^{d~*}}{M_{12}^d} }
          \frac{\bar{A}}{A},
\ee
where 
$A = {\cal A}(B^0 \to \phi K^0)$,
$\bar{A} = {\cal A}(\bar{B}^0 \to \phi \bar{K}^0)$ and
$M_{12}^d$ is given in Eq. (18).
The time-dependent CP asymmetry in $B \to \phi K$ decays are 
written in terms of $\lambda$ as
\be
a_{\phi K}(t) &\equiv& 
\frac{\Gamma(\bar{B}^0(t) \to \phi \bar{K}^0)
                    -\Gamma(B^0(t) \to \phi K^0)}
{\Gamma(\bar{B}^0(t) \to \phi \bar{K}^0)
                    +\Gamma(B^0(t) \to \phi K^0)},
\nonumber \\
&=& S_{\phi K} \sin \Delta m_B t - C_{\phi K} \cos \Delta m_B t,
\ee
where the coefficients 
\be
S_{\phi K} &=& \frac{2 {\rm Im} \lambda}{1+|\lambda|^2}, 
\nonumber \\ 
C_{\phi K} &=& \frac{1-|\lambda|^2}{1+|\lambda|^2} = -A_{\phi K},
\ee
are measured in the Belle and BaBar, of which average values are 
$- \eta S_{\phi K} = 0.44^{+0.17}_{-0.18}$, and
$C_{\phi K} = -0.23 \pm 0.15$, \cite{hfag}. 

\begin{figure}[t!]
\centering
%\hbox to\textwidth{\hss\epsfig{file=fig3.eps,width=10cm,height=10cm}\hss}
% \vskip -0.5cm
\includegraphics[width=10cm]{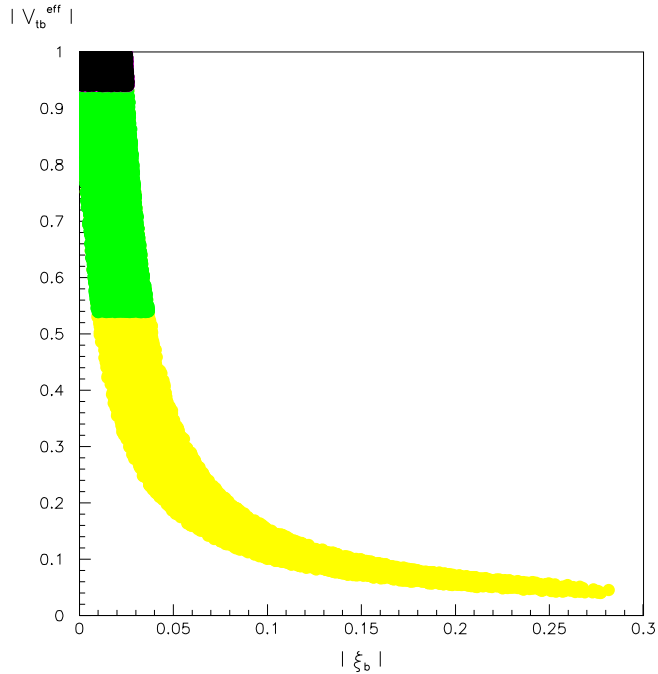}
\caption{\small
Allowed parameters $(|\xi_b|,|V_{tb}^{\rm eff}|)$ under the $B$ physics 
constraints and D0 dimuon asymmetry.
The whole band of the black + green (grey) + yellow (light grey) 
regions is allowed by Br$(B \to X_s \gamma)$ only.
The black + green (grey) regions are allowed by
Br$(B \to X_s \gamma)$, $\Delta M_s$ and $\Delta M_d$.
The black region is allowed by all constraints of
Br$(B \to X_s \gamma)$, $\Delta M_s$, $\Delta M_d$,
$S_{\phi K}$, $C_{\phi K}$, and
satisfies $A_{sl}^b$ measured by D0.
The confidence level is at 95 \% C.L..
}
\end{figure}
\begin{figure}[t!]
\centering
%\hbox to\textwidth{\hss\epsfig{file=fig4.eps,width=10cm,height=10cm}\hss}
% \vskip -0.5cm
\includegraphics[width=10cm]{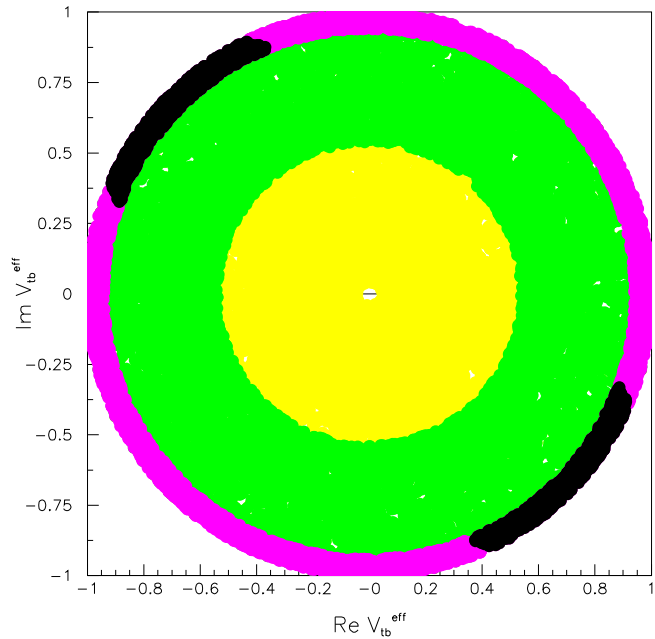}
\caption{\small
Allowed parameters $({\rm Re} V_{tb}^{\rm eff}, {\rm Im} V_{tb}^{\rm eff})$ 
constraints and D0 dimuon asymmetry.
The whole circle of the yellow (light grey) + green (grey) 
+ magenta (dark grey) + black regions is allowed 
by Br$(B \to X_s \gamma)$ only.
The thick ring of the green (grey) + magenta (dark grey) 
+ black regions allowed 
by Br$(B \to X_s \gamma)$ and $\Delta M_s$,
and the thin ring of the magenta (dark grey) + black regions allowed 
by Br$(B \to X_s \gamma)$, $\Delta M_s$, and $\Delta M_d$.
The black region is allowed by all constraints of
Br$(B \to X_s \gamma)$, $\Delta M_s$, $\Delta M_d$, 
$S_{\phi K}$, $C_{\phi K}$, and
satisfies $A_{sl}^b$ measured by D0.
The confidence level is at 95 \% C.L..
}
\end{figure}

\section{Results}

First we consider the nonzero anomalous $tsW$ couplings.
The $\bd$ mixing is not affected in this case and we get constraints
on the $tsW$ couplings from the $B \to X_s \gamma$ decay, $\Delta M_s$,
and CP asymmetry in $B \to \phi K$ decays.
Figure 1 shows the allowed parameters of $|\xi_s|$ and $|V_{ts}^{\rm eff}|$
at 95 \% C.L..
In the $B \to X_s \gamma$ decays of Eq. (17), 
the contribution of the right-handed couplings involves 
the enhancement factor $m_t/m_b$
and leads to substantial change of the amplitude.
Since the measurements of Br($B \to X_s \gamma$) agree with the SM
predictions, the substantial change of the amplitude due to $\xi_s$
should be compensated by a large shift of $V_{ts}^{\rm eff}$ 
as we can see in Fig. 1.
On the other hand, the contribution of $\xi_s$ to $M_{12}^s$ 
does not involve such an enhancement factor and
$M_{12}^s$ is governed merely by $V_{ts}^{\rm eff}$.
The like-sign dimuon charge asymmetry is affected through $M_{12}^s$.
Thus we find that the deviation of $A_{sl}^b$ from the SM value
leads to the deviation of $V_{ts}^{\rm eff}$ and also the nonzero $\xi_s$.
Finally these values satisfy 
the CP asymmetry in $B \to \phi K$ decays in most region.
We have allowed values of $V_{ts}^{\rm eff}$ and $\xi_s$
\be
0.01 < |\xi_s| < 0.03,~~~0.022 < |V_{ts}^{\rm eff}| < 0.029,
\ee
from all experimental constraints.
We find our results show sizable deviation from 
the value of $|V_{ts}| = 0.0403$ from the global fit
of the unitary triangle in the SM \cite{pdg}. 
Note that this result does not mean the violation of the CKM unitarity
but that an ``effective'' parameter $V_{ts}^{\rm eff}$ 
extracted from $B_s-\bar{B}_s$ mixing
looks different from the SM value. 

We show the allowed region of the complex parameter 
$V_{ts}^{\rm eff}$ at 95 \% C.L. in Fig. 2.
The sizable phase is predicted,
$14^o < \theta_{ts}^{\rm eff} < 22^o$ and  
$194^o < \theta_{ts}^{\rm eff} < 202^o$ 
from the measured $A_{sl}^b$ value in this plot,
while it is very small $\sim 2^o $ in the SM.
Note that this phase is essential to explain the dimuon charge asymmetry.
Since new effects on $\Gamma^q_{12}$ are ignored in this work,
our CP phase $\phi_s = -2 \theta_{ts}^{\rm eff}$
comes only from the $B_s-\bar{B}_s$ mixing.
Our results are consistent with the 2010 results 
$\phi_s({\rm CDF}) = (-29^{+44}_{-49})^o$ \cite{jpsiphi_cdf} and
$\phi_s({\rm D0}) = (-44^{+59}_{-51})^o$ \cite{jpsiphi_d0}
from $B_s \to J/\psi \phi$ decays
and also consistent with the recent best-fit value
$\phi_s = (-52^{+32}_{-25})^o$ at 2-$\sigma$ level
\cite{renz2}.
Such agreements are understood by that
all observed CP asymmetries at present in the $B_s$ system 
can be explained by the indirect CP violation through
modified $B_s-\bar{B}_s$ mixing.
%Recently reported are hints for deviations
%of experimantal data from the SM predictions in the $B_s$ system.
In our case, the modified $B_s$ mixing is due to $V_{ts}^{\rm eff}$.

Considering the anomalous $tbW$ couplings to explain $A_{sl}^b$,
we have constraints from $B \to X_s \gamma$ decay, $\Delta M_s$,
$\Delta M_b$, and the CP asymmetry in $B \to \phi K$ decays.
In Fig. 3, we show the allowed parameters of $|\xi_b|$ and 
$|V_{tb}^{\rm eff}|$ at 95 \% C.L..
In this case, the SM value of $|V_{tb}^{\rm eff}|=1$ is still
consistent with the dimuon charge asymmetry.
Instead we require new phase of $V_{tb}^{\rm eff}$ 
to explain the $A_{sl}^b$ as shown Fig. 4
although $V_{tb}$ is real in the SM.
We used the SM value of the CP violating phase 
$\phi_d^{\rm SM} = -0.091^{+0.026}_{-0.038}$ \cite{nierste}.
Figure 4 allows the phase angle $-66 ^o < \theta_{tb} < -21^o$ 
and $114 ^o < \theta_{tb} < 159^o$ 
at 95 \% C.L..
However, the CP phase of $B_d$ system is precisely measured 
in $B \to J/\psi K_s$ and the recent world average value is given by 
\cite{hfag}
$\sin 2 \beta = 0.676 \pm 0.020$,
which agrees with the SM predictions very well.
Then the large additional phase of $V_{tb}$ 
is not consistent with the measured $\sin 2 \beta$.
Such disagreement implies that it is hard to explain
the dimuon charge asymmetry and the $B \to J/\psi K$ decay simultaneously
only with the modification of $V_{tb}^{\rm eff}$. 
Thus we conclude that the dimuon charge asymmetry favours
the anomalous $tsW$ couplings rather than $tbW$ couplings.

Since the anomalous $tsW$ couplings contribute to $M_{12}^s$ 
and not to $M_{12}^d$,
only $a_{sl}^s$ is shifted as $\xi_s$ varies.
Meanwhile, both $M_{12}^s$ and $M_{12}^d$ are affected 
by the anomalous $tbW$ couplings 
and also both $a_{sl}^s$ and $a_{sl}^b$ are modified
as $\xi_b$ varies.
We show the variation of $a_{sl}^s$ and $a_{sl}^b$ in Fig. 5
with the allowed parameter sets of $(\xi_s,V_{ts}^{\rm eff})$
and $(\xi_b,V_{tb}^{\rm eff})$ 
given in Fig. $1-4$.

\begin{figure}[t!]
\centering
%\hbox to\textwidth{\hss\epsfig{file=fig5.eps,width=10cm,height=10cm}\hss}
% \vskip -0.5cm
\includegraphics[width=10cm]{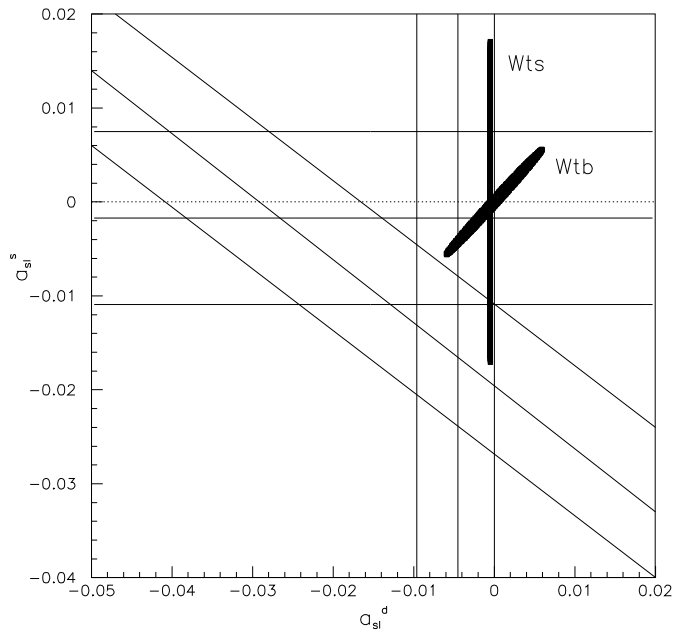}
\caption{\small
The thick black lines are our predictions of $a_{sl}^d$ and $a_{sl}^s$
varying the anomalous $tbW$ and $tsW$ couplings 
with the measurements of $A_{sl}^b$ (inclined band) by D0 \cite{d0new},
$a_{sl}^d$ (vertical band) at $B$ factory \cite{hfag} 
and $a_{sl}^s$ (horizontal band) by D0 \cite{d0old}.
The crossing point of thick lines denotes the SM prediction.
The $1-\sigma$ error bands are shown.
}
\end{figure}

\section{Concluding Remarks}

We have studied the effects of the anomalous $tsW$ and $tbW$ couplings
to explain the recently measured deviation of 
like-sign dimuon charge asymmetry at Tevatron.
Our new complex couplings 
are able to explain the D0 dimuon charge asymmetry at 95 \% C.L.
under constraints from the precisely measured Br($B \to X_s \gamma$), 
$\Delta M_d$, $\Delta M_s$, $S_{\phi K}$, and $C_{\phi K}$ data.
However the additional phase of $V_{tb}^{\rm eff}$ is not
consistent with the CP violation in $B \to J/\psi K$ decay,
while the anomalous $tsW$ couplings agree with that 
in $B \to J/\psi \phi$ decays at 2-$\sigma$ level.
We conclude that the dimuon charge asymmetry favours a new top couplings
in $B_s-\bar{B}_s$ mixing than in $B_d-\bar{B}_d$ mixing,
and show that the anomalous $tsW$ couplings
satisfies constraints of $B$ physics.

\acknowledgments
This work was supported by
the Basic Science Research Program through the National Research Foundation
of Korea (NRF) funded by the Korean Ministry of 
Education, Science and Technology (2009-0088395).
KYL is supported in part by WCU program through the KOSEF funded 
by the MEST (R31-2008-000-10057-0)
and the Basic Science Research Program 
through the National Research Foundation
of Korea (NRF) funded by the Korean Ministry of 
Education, Science and Technology (2009-0076208).

%%%%%%%%%%%%%%%%%% References
%%%%%%%%%%%%%%%%%%%%%%%%%%%%%%%%%%%%%%%%%%%%%%%%%%%%%%%%%%%%%%%%%%%%%%%
\def\PRD #1 #2 #3 {Phys. Rev. D {\bf#1},\ #2 (#3)}
\def\PRL #1 #2 #3 {Phys. Rev. Lett. {\bf#1},\ #2 (#3)}
\def\PLB #1 #2 #3 {Phys. Lett. B {\bf#1},\ #2 (#3)}
\def\NPB #1 #2 #3 {Nucl. Phys. {\bf B#1},\ #2 (#3)}
\def\ZPC #1 #2 #3 {Z. Phys. C {\bf#1},\ #2 (#3)}
\def\EPJ #1 #2 #3 {Euro. Phys. J. C {\bf#1},\ #2 (#3)}
\def\JHEP #1 #2 #3 {JHEP {\bf#1},\ #2 (#3)}
\def\IJMP #1 #2 #3 {Int. J. Mod. Phys. A {\bf#1},\ #2 (#3)}
\def\MPL #1 #2 #3 {Mod. Phys. Lett. A {\bf#1},\ #2 (#3)}
\def\PTP #1 #2 #3 {Prog. Theor. Phys. {\bf#1},\ #2 (#3)}
\def\PR #1 #2 #3 {Phys. Rep. {\bf#1},\ #2 (#3)}
\def\RMP #1 #2 #3 {Rev. Mod. Phys. {\bf#1},\ #2 (#3)}
\def\PRold #1 #2 #3 {Phys. Rev. {\bf#1},\ #2 (#3)}
\def\IBID #1 #2 #3 {{\it ibid.} {\bf#1},\ #2 (#3)}
%%%%%%%%%%%%%%%%%%%%%%%%%%%%%%%%%%%%%%%%%%%%%%%%%%%%%%%%%%%%%%%%%%%%%%%

\end{document}